\documentclass[apj,graphicx]{emulateapj}
\usepackage{apjfonts}

\begin{document}

\title{Investigating the nature of absorption lines in the {\it Chandra} X-ray spectra of the neutron star binary 4U~1820$-$30}
\shortauthors{Cackett et al.}
\shorttitle{X-ray absorption lines in 4U~1820$-$30}

\author{E.~M.~Cackett\altaffilmark{1,2}, J.~M.~Miller\altaffilmark{1}, J.~Raymond\altaffilmark{3}, J.~Homan\altaffilmark{4}, M.~van~der~Klis\altaffilmark{5},\\ M.~M\'endez\altaffilmark{5,6,7},  D.~Steeghs\altaffilmark{3}, R.~Wijnands\altaffilmark{5}}
\email{ecackett@umich.edu}
\altaffiltext{1}{Department of Astronomy, University of Michigan, 500 Church St, Ann Arbor, MI 48109-1042, USA}
\altaffiltext{2}{Dean McLaughlin Postdoctoral Fellow}
\altaffiltext{3}{Harvard-Smithsonian Center for Astrophysics, 60 Garden Street, Cambridge, MA 02138, USA}
\altaffiltext{4}{MIT Kavli Institute for Astrophysics and Space Research, MIT, 70 Vassar Street, Cambridge, MA 02139-4307}
\altaffiltext{5}{Astronomical Institute `Anton Pannekoek', University of Amsterdam, Kruislaan 403, 1098 SJ, Amsterdam, the Netherlands}
\altaffiltext{6}{SRON - Netherlands Institute for Space Research, Sorbonnelaan 2, 3584 CA Utrecht, the Netherlands}
\altaffiltext{7}{Astronomical Institute, University of Utrecht, PO Box 80000, 3508 TA Utrecht, The Netherlands}

\begin{abstract}
We use four {\it Chandra} gratings spectra of the neutron star low-mass X-ray binary 4U~1820$-$30 to better understand the nature of certain X-ray absorption lines in X-ray binaries, including the \ion{Ne}{2}, \ion{Ne}{3}, \ion{Ne}{9}, \ion{O}{7}, and \ion{O}{8} lines.  The equivalent widths of the lines are generally consistent between the observations, as expected if these lines originate in the hot interstellar medium.  No evidence was found that the lines were blueshifted, again supporting the interstellar medium origin, though this may be due to poor statistics.  There is apparent variability in the \ion{O}{8} Ly$\alpha$ line equivalent width providing some evidence that at least some of the \ion{O}{8} absorption arises within the system.  However, the significance is marginal (2.4-$\sigma$), and the lack of variation in the other lines casts some doubt on the reality of the variability.  From calculating the equivalent hydrogen column densities for a range of Doppler parameters, we find they are consistent with the interstellar origin of the lines. Additionally we fit the spectra with photoionization models for locally absorbing material, and find that they can reproduce the spectrum well, but only when there is an extremely low filling factor.  We conclude that both the ISM and local absorption remain possible for the origin of the lines, but that more sensitive observations are needed to search for low-level variability.
\end{abstract}


\section{Introduction}
High-resolution X-ray spectroscopy from {\it Chandra} and {\it XMM-Newton} allows for the study of absorption and emission lines in Galactic X-ray binaries.  While emission lines are clearly associated with the sources themselves \citep[e.g.,][]{cottam01_0748,cottam01_1822,schulz01}, that is not always the case for absorption lines.  In some systems, the absorption lines are also clearly associated with the source, as they are observed to be variable and/or blueshifted \citep[e.g.,][]{lee02,parmar02,ueda04,boirin05,diaztrigo06,miller06_nature,miller06_H1743}.  However, in other sources the lines are interpreted as being associated with absorption by the hot ionized interstellar medium, which is thought to be present in the disk and halo of the galaxy, with temperatures around $10^6$ K \citep[e.g.,][and references therein]{futamoto04,yao_wang05,juett06}.

From studies of the absorption lines in the black hole X-ray binary GRO~J1655$-$40, the plethora of highly-ionized lines that are detected are all seen to be blueshifted -- evidence for a mass outflow into our line of sight, presumably from the accretion disk.  In fact, detailed modeling of the lines in this system suggest that magnetic fields may be the force driving the wind \citep{miller06_nature}.   Other black hole systems also show evidence for highly-ionized outflows such as H~1743$-$322, GX~339$-$4 and GRS~1915+105 \citep{miller06_H1743,miller04_gx339,lee02}.  Disk winds are also observed from the accretion disks around supermassive black holes in Active Galactic Nuclei \citep[e.g.][]{crenshaw99,kaastra00,kaspi00}.
 
Several neutron star low-mass X-ray binaries also display clear evidence for absorption local to the source.  For example, GX 13+1 shows blueshifted absorption lines indicating an outflow with a velocity of $\sim$400 km s$^{-1}$ \citep{ueda04}.  Moreover, in the X-ray dipping sources, strong absorption lines are observed as these objects are inclined close to the orbital plane  \citep[e.g.,][]{sidoli01,parmar02,boirin05,church05,diaztrigo06}.  On the other hand, the nature of the absorption lines observed in many other neutron star X-ray binaries is less clear - the lines are weaker and there is no evidence of significant blueshifts, supporting the interpretation that these lines are due to the interstellar medium \citep[e.g.,][]{yao_wang05,juett06}.

As part of the {\it Chandra} HETGS Z/Atoll Spectroscopic Survey (CHAZSS) we have obtained sensitive observations of 6 neutron star low-mass X-ray binaries at 2 separate epochs using the {\it Chandra} high energy transmission grating spectrometer (HETGS). The multiple epochs allow for a search for variability in the absorption lines present in these systems; this is key for learning about the nature of the absorption lines.  In this paper, we present the observations of the system 4U~1820$-$30 from this survey.  Tight limits can be set on the size of any absorbing region associated with a binary if the system is ultra-compact in nature.  Therefore, the ultra-compact nature of 4U~1820$-$30 \citep*{stella87} makes it a particularly interesting source to study absorption lines.

\section{Previous {\it Chandra} Observations of 4U~1820$-$30}
4U~1820$-$30 is an accreting neutron star low mass X-ray binary in the globular cluster NGC~6624, which has a well determined distance \citep[$7.6 \pm 0.4$ kpc][]{kuulkers03} and reddening \citep[{\it E(B-V)} = $0.32 \pm 0.03$][]{kuulkers03,bohlin78}.  This reddening is consistent with the total HI column density of our Galaxy in this direction determined from 21 cm radio observations, $N_H = 1.5 \times 10^{21}$ cm$^{-2}$ \citep{dickeylockman90}.  We adopt the above column density throughout this paper.

Absorption lines and edges in the X-ray spectrum of this source have been studied previously by \citet{futamoto04}, \citet{yao_wang05} and \citet{yao_wang06} using an LETG/HRC {\it Chandra} observation (\dataset[ADS/Sa.CXO#obs/00098]{ObsId 98}), and by \citet{juett04}, \citet{yao_wang05}, \citet{juett06} and \citet{yao_wang06} using HETG/ACIS {\it Chandra} observations (\dataset[ADS/Sa.CXO#obs/01021]{ObsIds 1021} and \dataset[ADS/Sa.CXO#obs/01022]{1022}).  A summary of the {\it Chandra} observations of this source is given in Table \ref{tab:obs}.  We briefly summarize the main results of these previous studies below.

\citet{futamoto04} clearly detect the \ion{O}{7} He$\alpha$, \ion{O}{8} Ly$\alpha$, and \ion{Ne}{9} He$\alpha$ absorption lines in the LETG spectrum, and put upper limits on the \ion{O}{7} He$\beta$ line.  From their curve of growth analysis and photoionization modeling, they deduce that all oxygen will be fully photoionized if the absorbing column is located close to the binary system, and therefore attribute these lines to hot gas in the ISM.  However, we note that they make two tacit assumptions - that the filling factor of the gas is 1, and that all the absorption arises from the same gas.

The studies by \citet{juett04} and \citet{juett06} are concerned with the absorption edges from oxygen, neon and iron, comparing the spectra of a number of X-ray binaries.  In \citet{juett04} the structure around the oxygen edge in 7 X-ray binaries is studied, looking at the \ion{O}{1}, \ion{O}{2}, and \ion{O}{3} absorption lines in that region.  The edge and these low ionization lines are attributable to ISM absorption.  In \citet{juett06} the neon and iron L edges are studied, and the nature of the neon absorption lines discussed.  These authors come to the conclusion that the neon lines in the low-mass X-ray binaries studied are consistent with predictions for the ISM, except for the \ion{Ne}{9} line detected in the black hole X-ray binary GX~339$-$4 \citet{miller04_gx339} which shows a column density significantly higher than predicted for the ISM, confirming that the largest contribution to this line (in this source, at least) is from local absorption.

From a study of 10 low-mass X-ray binaries, \citet{yao_wang05} find that the detected \ion{Ne}{9}, \ion{O}{7} and \ion{O}{8} absorption lines are consistent with the hot ISM origin.  \citet{yao_wang06} focuses specifically on 4U~1820$-$30, discussing the implications of the Ne and O abundance in the ISM from the absorption lines/edges in the {\it Chandra} observations.

\section{Observations and Data Reduction}
All the previous observations discussed above suffer from pileup, where a high count rate leads to more than one photon arriving at a pixel per read out.  This affects the shape of the spectrum, typically making it harder as the sum of the energies of the multiple photons that arrived between each readout is detected as one count.  It is unclear how pileup affects the measurement of absorption lines.  It likely modifies the continuum level and shape, which would in turn change any measured equivalent widths (which depends on the continuum flux across the line).  Currently, there is no unambiguous way to model pileup in {\it Chandra} gratings spectra.  To minimize pileup we operated {\it Chandra} with the High Energy Transmission Grating Spectrometer in continuous clocking mode to significantly increase the temporal resolution.  Continuous clocking mode allows milli-second timing at the expense of one dimension of spatial information.

As part of CHAZSS, 4U~1820$-$30 was observed twice with {\it Chandra} for approximately 25 ksec each time, first on 2006 Aug 12 (\dataset[ADS/Sa.CXO#obs/06633]{ObsId 6633}), and later on 2006 Oct 20 (\dataset[ADS/Sa.CXO#obs/06634]{ObsId 6634}).  The HETGS-dispersed spectrum was read-out with the ACIS-S array operating in continuous-clocking mode.  We reduced these data using CIAO version 3.3.0.1 and following the standard analysis threads\footnote{http://http://cxc.harvard.edu/ciao/threads}.  Using the {\it tgextract} tool spectra the +1 and -1 order were extracted at the nominal instrument resolution for both the high-energy grating (HEG) and the medium-energy grating (MEG).  Using the standard redistribution matrix files (RMFs) from the CALDB, the ancillary response files (ARFs) were generated using the {\it fullgarf} script.  We carefully inspected the MEG +1 and -1 orders to confirm there was no wavelength shift between the two orders. The spectra and ARFs were then added together using the {\it add\_grating\_spectra}, giving a first-order spectrum for the MEG.  Similarly, we created a first-order spectrum for the HEG.

In addition to analyzing these two new {\it Chandra} observations, we also re-analyze the two previous {\it Chandra} HETGS observations of 4U~1820$-$30 (ObsId 1021 \& 1022) to allow for a search for variability in the absorption lines.  These two datasets were not operated in continuous-clocking mode, and suffer from pile-up.  We reduce the data following the standard analysis threads, extracting the spectra and ARFs using the same method described above.
For consistency, we only analysis the HETGS observations here and not the LETGS observation which has a lower spectral resolution.

\section{Analysis and Results}
\subsection{Absorption lines}

When searching for absorption lines, we use only the MEG spectra, as the MEG has a higher effective area over the region of interest compared to the HEG.  We inspected the spectra over 2\AA~segments from 2\AA ~to 24\AA ~ using the ISIS spectral fitting package.  In each segment, we fitted a simple power-law (modified, where appropriate, by an absorption edge due to neutral elements in the ISM). Within each segment, we searched for any significant absorption lines present.  We fitted a Gaussian to the detected absorption lines  whose central wavelength, width, and normalization were all allowed to be free parameters.  The equivalent width of the line was determined from the fitted Gaussian.  For line identifications we use \citet*{verner96} as a reference.  However, for \ion{Ne}{2} and \ion{Ne}{3} we use the wavelengths determined by \citet{juett06}.  They find the weighted mean of the central wavelengths of these lines measured in multiple X-ray binaries.  The wavelengths for these lines from different sources are seen to consistent with each other, and are all systematically shifted by a small amount ($\sim$20m\AA) from the theoretical value determined by \citet{behar_netzer02}, though within their stated errors.  The \citet{juett06} values are known most accurately, thus we use them throughout the rest of the paper.

In the two new {\it Chandra} observations, we detect several X-ray absorption lines.  We define a detected line as one for which fits with a Gaussian give an equivalent width that excludes zero at more than the 2$\sigma$ level of confidence (note that we are being conservative by using the significance with which we can measure the equivalent width as the significance of the detection).  In increasing wavelength (theoretical/best observational wavelengths are given in brackets), the detected lines are: \ion{Ne}{9} (13.447 \AA), \ion{Ne}{3} (14.508 \AA), \ion{Ne}{2} (14.608 \AA), \ion{O}{7} (18.629 \AA), \ion{O}{8} (18.967 \AA).  All these absorption lines are detected in both observations, however, some were not detected in the previous {\it Chandra} observations, likely due to the lower sensitivity of those shorter observations.  Whilst the \ion{O}{1} absorption line at $\sim$23.5\AA~was detected in all observations, we have not attempted to fit this line due to the complicated structure around this region due to multiple oxygen lines.  The structure in this region has been studied in detail previously by \citet{juett04}.  The interstellar absorption at this wavelength also significantly increases the noise in the data, therefore the LETG observation of 4U~1820$-$30 previously studied present the best available dataset for analysis of this region.  We also note that the \ion{O}{7} line at $\sim$21.6\AA~was not detected in any of these observations, due to the low signal-to-noise in that region.  Again, the previous LETG observation of 4U~1820$-$30 gives the best detection of this line.

\begin{figure}
\centering
 \begin{tabular}{lr}
\includegraphics[width=8cm]{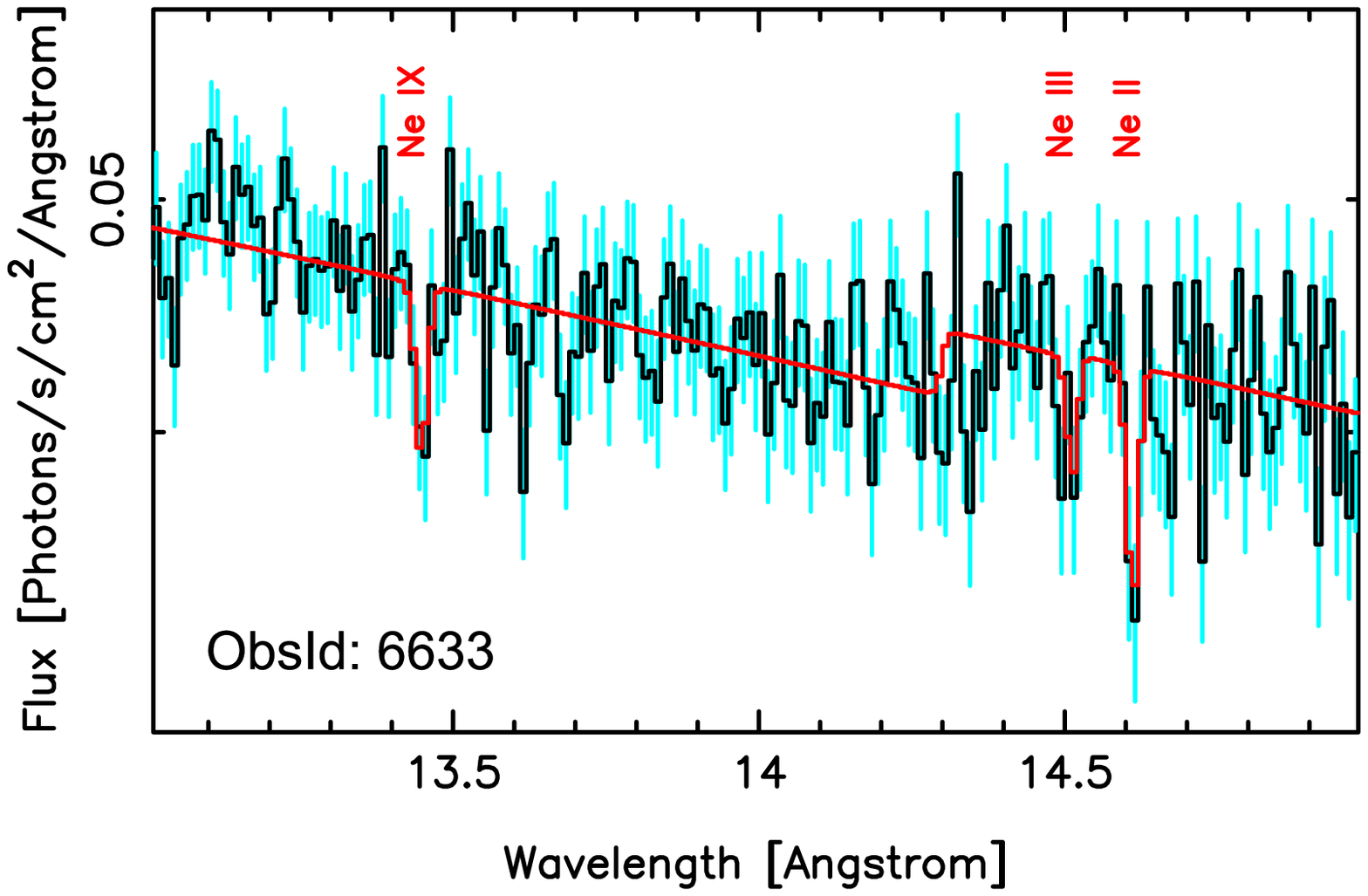} \hspace{0.2cm}
\vspace{0.3cm}
\hspace{-0.5cm}

&
\includegraphics[width=8cm]{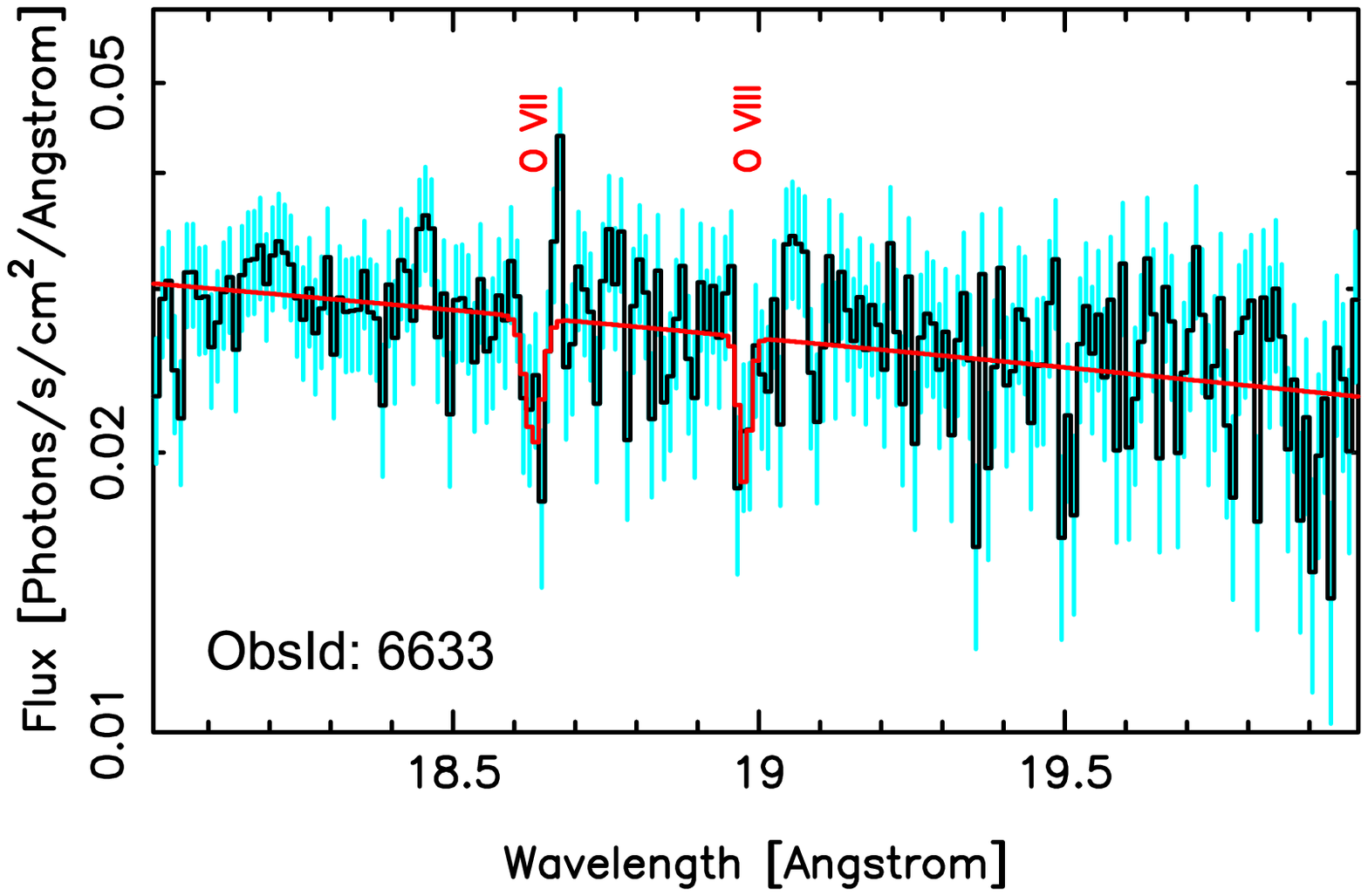} \\
\includegraphics[width=8cm]{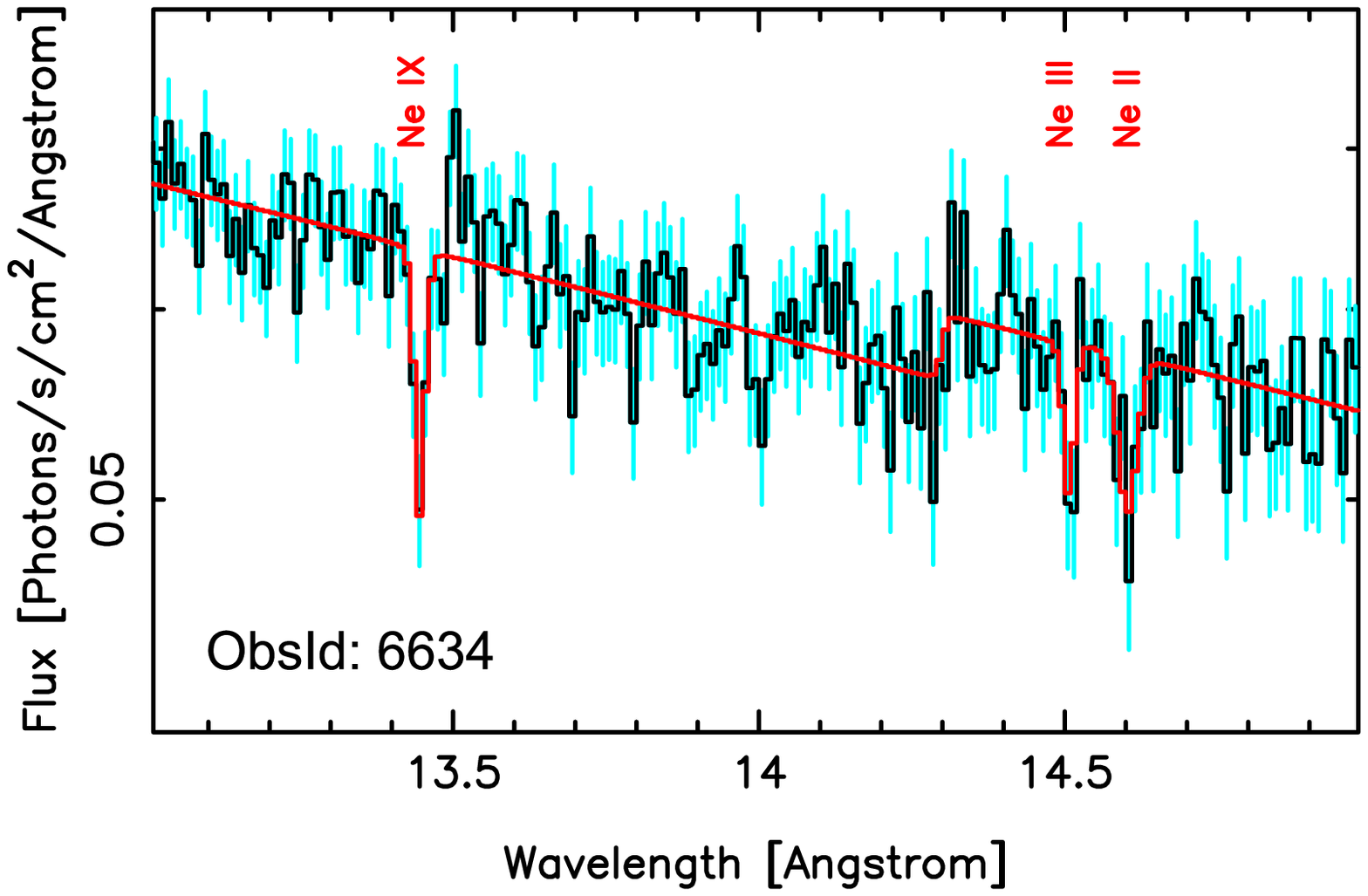}
&
\includegraphics[width=8cm]{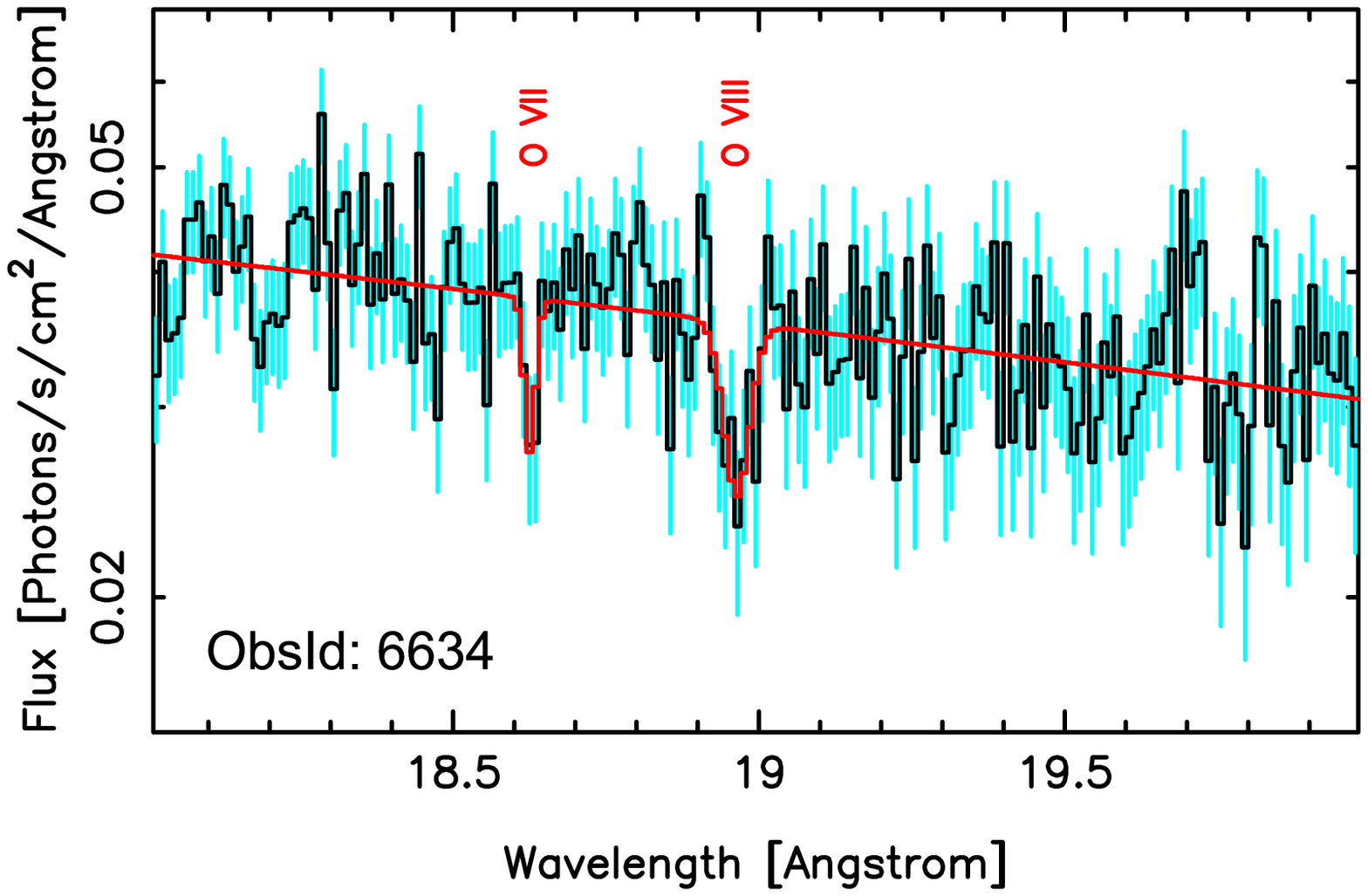}
\end{tabular}
\caption{$13 - 15$\AA~and $18 - 20$\AA~segments of the combined first-order MEG spectra of 4U~1820$-$30.  For illustrative purposes the data have been rebinned by a factor of 2.  The top panels show these regions in ObsId 6633, and the bottom panels show these regions in ObsId 6634.  The data is in black, and the 1$\sigma$ error bars are in blue.  The best-fitting model is in red.  The continuum is fit by a power-law modified by neutral ISM absorption edges, where appropriate.  Absorption lines are modeled by a Gaussian.  Table \ref{tab:lines} gives the line properties.}
\label{fig:lines}
\end{figure}

In cases where an absorption line had been detected in one of the other observations, but not in the observation we were studying, a 95\% confidence level upper limit for the line flux and equivalent width was determined.  This was achieved by fixing the line wavelength at the theoretical value, fixing the width at a value smaller than the instrument resolution (as would be the case if the line was unresolved), and then increasing the line flux until the $\chi^2$ value for the fit increased by 2.7.  We justify fixing the width smaller than the instrument resolution as for the MEG, the resolution corresponds to a FWHM of $\sim$350 km s$^{-1}$ at 18.967\AA, and higher at shorter wavelengths. This is significantly higher than the expected thermal broadening in the ISM.  Table \ref{tab:lines} gives the line properties determined from the spectral fitting.  In Fig. \ref{fig:lines} we show the absorption lines detected in the two new {\it Chandra} observations.

As \ion{Ne}{9} is detected at 13.447\AA~ one should also expect to find additional \ion{Ne}{9} lines at 11.547\AA~ and 11.000\AA.  However, we do not clearly detect these lines in any of the observations.  If the line is not saturated, the ratio of the equivalent widths should follow the ratio of the oscillator strengths times the ratio of the wavelengths squared.  The mean value of the equivalent width of the \ion{Ne}{9} (13.447\AA) line is approximately 5m\AA.  From the oscillator strengths, this would put the expected equivalent widths for the \ion{Ne}{9} lines at 11.547\AA~ and 11.000\AA~ to be 0.8m\AA~ and 0.3m\AA, respectively.  In ObsId 6633, we note that there may be a weak absorption line present at $\lambda = 11.550 \pm 0.007$ \AA~, with an equivalent width of EW = $1.8 \pm 0.9$ m\AA, though this is only a 2$\sigma$ detection.  This measured equivalent width is consistent with what is predicted from the oscillator strengths.  The line at 11.000\AA~ is not detected, and we determine the 95\% confidence level upper limit to be 1.7 m\AA~ - the line would be too weak to detect given the noise in the data.  In ObsId 6634 these \ion{Ne}{9} lines are not detected, and we find the upper limits on the equivalent width for the lines at 11.547\AA~ and 11.000\AA~ to be 1.6m\AA~ and 1.3m\AA~ respectively, consistent with what would be expected from the oscillator strengths.  We do not determine upper limits for these lines from the ObsId 1021 and 1022 datasets, as they are substantially less sensitive, and therefore the lines could not be detected at the predicted equivalent widths.  While the upper limits do not rule out that the line is saturated, we cannot be conclusive.

Similarly, we try to determine whether the \ion{O}{8} lines are saturated by determining limits on the \ion{O}{8} Ly$\beta$ line at 16.006\AA.  Fitting a Gaussian, with central wavelength fixed at 16.006\AA~to ObsId 6633 and 6634 we measure equivalent widths of $3.0\pm1.5$ m\AA, and $3.4\pm2.8$ m\AA~respectively.  This gives the ratio of the equivalent widths of the \ion{O}{8} Ly$\beta$ to Ly$\alpha$ as $0.47\pm0.32$ and $0.18\pm0.16$ for ObsId 6633 and 6634 respectively.  For comparison, if the lines are not saturated we would expect this ratio to be 0.14 given the oscillator strengths and wavelengths.  The measured value from ObsId 6634 is clearly consistent with this ratio, and in ObsId 6633 the ratio is not significantly different.  Therefore the \ion{O}{8} lines are not highly saturated.

Another check on saturation can be performed using the measured equivalent width of the \ion{O}{7} Ly$\alpha$ line \citep[measured in previous {\it Chandra} LETG observations by][]{futamoto04} and comparing it with the weighted average of the \ion{O}{7} Ly$\beta$ equivalent widths measured here.  We find that the ratio of \ion{O}{7} Ly$\alpha$ equivalent width to Ly$\beta$ equivalent width is $5.5\pm2.2$.  From the line wavelengths and oscillator strengths \citep[taken from][]{verner96}, the expected ratio in the non-saturated regime is 6.4, consistent the ratio from the observed equivalent widths, suggesting that the \ion{O}{7} lines are not highly saturated.

\subsection{Column Densities}

From the measured equivalent widths of the lines, we determine the column densities.  Firstly we assume that the lines are not saturated and on the linear part of the curve of growth.  Under this assumption, the column density and equivalent width are associated via $W_{\lambda} = (\pi e^2 / M_e c^2) N_j \lambda^2 f_{ij} = 8.85\times10^{-13} N_j \lambda^2 f_{ij}$, where $W_\lambda$ is the equivalent width (in cm), $N_j$ is the column density of a given species, $\lambda$ is the line wavelength (in cm) and $f_{ij}$ is the oscillator strength \citep{spitzer78}. 

However, as we cannot conclusively determine whether the lines are saturated, we also calculate the column density for a range of Doppler parameters, $b$.  For a lower limit for the line broadening, we estimate the FWHM of the Galactic rotation in the direction of the source from a Besancon model of the Galaxy to be 50 km s$^{-1}$.  We also evaluate the column densities assuming $b = 100$ km s$^{-1}$ and $b = 200$ km s$^{-1}$.  The latter being an upper limit on the width of \ion{O}{6} emission lines from UV observations with {\it FUSE} \citep{otte06}.  Given the oscillator strength, line wavelength, observed equivalent width and assumed Doppler parameter, we determine the required column density for each line in ObsIds 6633 and 6634, using the curve of growth equations in \citet{spitzer78}.For the blended \ion{O}{8} fine structure components we compute the column densities numerically.  Table \ref{tab:densities} gives the calculated column densities.

In order to test whether these column densities are consistent with the interstellar medium, we converted the column density for each species into an equivalent hydrogen column density.  To do this we require the ionic fraction for each ion, which for collisional ionization (the mechanism present in the ISM) depends on the temperature.  The low ionization lines (\ion{Ne}{2} and \ion{Ne}{3}) will be in a lower temperature gas than the higher ionization lines (\ion{Ne}{9}, \ion{O}{7}, \ion{O}{8}), thus we estimated the temperatures separately.  Comparing the observed ratio of \ion{Ne}{2}/\ion{Ne}{3} with the ratio of the ionic fractions for those ions at a range of temperatures \citep[as calculated for collisional ionization by][]{mazzotta98}, we determined the best-fitting temperature of the gas.  Given that temperature, the ionic fractions at that temperature and the ISM abundances from \citet{wilms00} we converted the ionic column densities into equivalent hydrogen column densities (see Table \ref{tab:densities}).  We found the temperatures for the \ion{Ne}{2} and \ion{Ne}{3} gas to be in the range $(4.8-5.2)\times10^4$ K.

Similarly for \ion{Ne}{9}, \ion{O}{7}, \ion{O}{8} we determined a temperature from the ratios of \ion{O}{7}/\ion{O}{8} and \ion{O}{7}/\ion{Ne}{9} \citep[assuming Ne/O from][]{wilms00} compared to the ionic fractions from \citet{mazzotta98}, and then used the estimated ionic fractions and abundances to convert to equivalent hydrogen column densities (see Table \ref{tab:densities}).  The temperatures we determine for this gas is in the range $(1.3-2.0)\times10^6$ K, consistent with the hot phase of the ISM.

The equivalent hydrogen column densities that we estimate are all consistent with having a lower column density than the column to the source as seen in \ion{H}{1} \citep[$1.5\times10^{21}$ cm$^{-1}$,][]{dickeylockman90}.

\subsection{Photoionization modeling}
In order to assess the possibility of the observed absorption lines originating from local absorption in the X-ray binary system (rather than the ISM) we used the photoionization code XSTAR v2.1 \citep{kallman01} to model the observed spectrum in ObsID 6633 and 6634.  For the input ionizing spectrum we use the unabsorbed source spectra for each observation, scaled to the observed luminosities (see section \ref{sec:flux} for details on determining the source flux, we assume $d=7.6$ kpc giving 0.5-10 keV luminosities of $5.7\times10^{37}$ and $8.1\times10^{37}$ erg s$^{-1}$ for ObsID 6633 and 6634, respectively).  For each observation we then created a grid of absorption line models allowing the column density and ionization parameter, $\xi = L/nR^2$, to vary.  In our models we assume solar abundances, a covering fraction of 0.5, and turbulent velocities of 200 km s$^{-1}$.  From these grids of models we then fit the observed spectra (again using ISIS) from 13\AA~to 20\AA~with a power-law (with Galactic absorption fixed at $1.5\times10^{21}$ cm$^{-1}$) convolved with the absorption line models calculated from XSTAR.  This allows us to determine under what conditions the observed spectrum can be reproduced by photoionization of local absorbing gas.

\begin{figure}
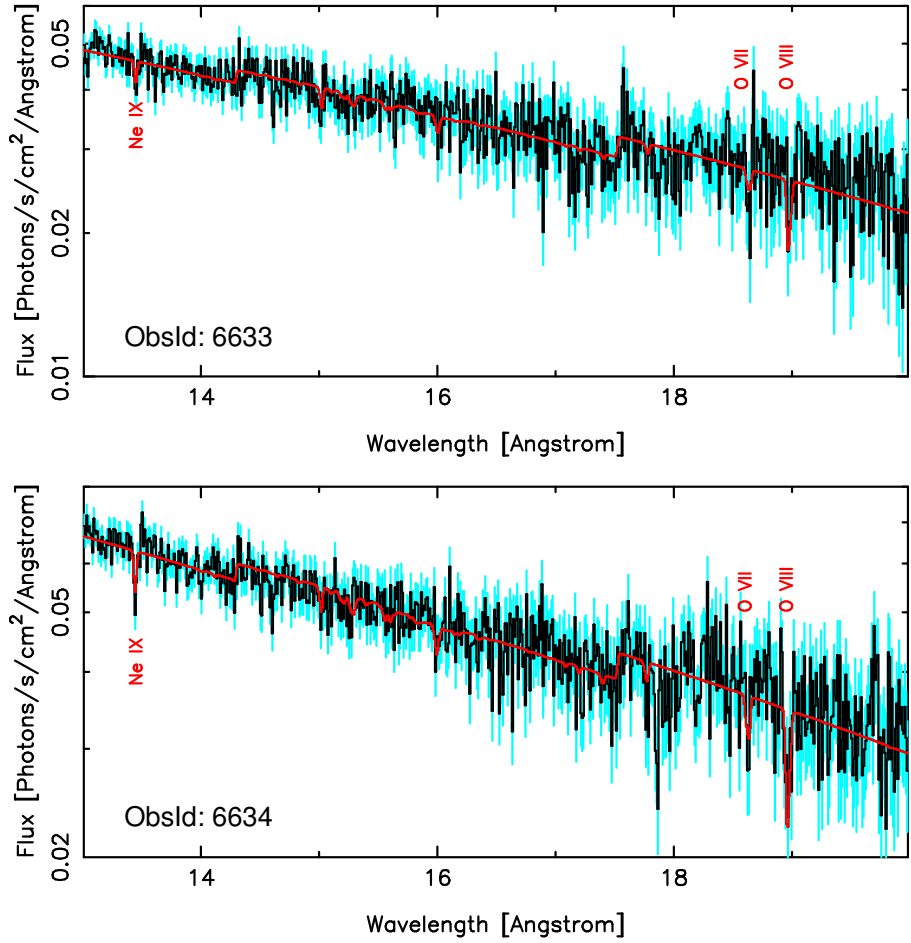

\centering
\includegraphics[width=12cm]{f2a.ps} \\
\vspace{0.3cm}
\includegraphics[width=12cm]{f2b.ps}
\caption{Best-fitting XSTAR photionization models for ObsID 6633 (top) and 6634 (bottom).  The \ion{Ne}{9}, \ion{O}{7} and \ion{O}{8} lines are reproduced well by the models.}
\label{fig:xstar_models}
\end{figure}

In Fig. \ref{fig:xstar_models} we show the best-fitting models from the photoionization modeling.  We find that for both observations, the observed \ion{Ne}{9}, \ion{O}{7}, \ion{O}{8} lines can be reproduced well with $\chi^2_\nu = 1.13$ for ObsID 6633, and $\chi^2_\nu = 1.15$ for ObsID 6634.  Note that the \ion{Ne}{2} and \ion{Ne}{3} lines are not included in the code, and thus are not reproduced.

For ObsID 6633 we find the best fit with $N_H = (6.1\pm1.5) \times 10^{19}$ cm$^{-2}$ and $\log{\xi}=1.36\pm0.05$ (where $\xi = L/nR^2$).  For ObsID 6634 we find the best fit with $N_H = (7.6 \pm 1.3) \times 10^{19}$ cm$^{-2}$ and $\log{\xi}=1.34 \pm 0.03$.  Given that the binary separation for 4U~1820$-$30 is $\sim10^{10}$ cm \citep{stella87}, with the ionization parameters we find the gas density would have to be $n = 2.5\times10^{16}$ cm$^{-3}$ for ObsID 6633 and $n = 4\times10^{16}$ cm$^{-3}$ for ObsID 6634 for the largest plausible radius for absorbing gas within the system.  The implies that the filling factor, $f$, is extremely low with $f = N_H/nR = 2.4\times10^{-7}$ for 6633 and $f=1.9\times10^{-7}$ for 6634 .  Since $f$ scales with $R$, smaller radii within the system make the problem worse.

It seems unlikely that the gas would be in a shell, as the filling factor would mean that the thickness of the shell would be very small (on the order of only tens of metres for gas within the system).   Additionally it is also unlikely to be a shell covering the entire 4$\pi$ or else one would also observe emission lines.  One possibility is that thermal instabilities produce small, dense structures, or alternatively there could be partial covering by larger blobs, which could be driven by a wind.  Therefore, while it seems possible that locally photoionized material could account for the absorption, it does require fairly extreme parameters.  We note that \citet{futamoto04} use CLOUDY to run a numerical simulation for previous observations of 4U~1820$-$30.  These authors concluded that photoionized gas within the binary system could not reproduce the observed line column densities.  In summary, the ISM may be the more likely source of the absorption lines, but our models demonstrate that photoionization cannot be ruled out, especially given the presence of variability.

\subsection{Source flux}\label{sec:flux}

In order to compare absorption line properties with the source properties, we determine the source flux from the best fitting continuum model to the HEG spectrum.  For this broadband spectral fitting we use XSPEC \citep{arnaud96}.  We do not fit the MEG and HEG simultaneously due to the significant pileup in the first two observations, where the observations were not operated in continuous clocking mode.  As the MEG has a larger effective area than the HEG at low energies (by about a factor of 2) it is more substantially affected by pileup.  This is clearly apparent when trying to fit a continuum model to just the MEG spectra in these observations.  There is an excess of counts at high energies and deficit at low energies, as expected by pileup.  The HEG is much less affected, and thus we choose to fit to just the HEG 1st order spectra.  For consistency, we use this method for fitting to the new {\it Chandra} observations also, even though they do not suffer from pileup.

We choose to fit an absorbed blackbody plus power-law model to the HEG spectra over the energy range $1.2 - 8$ keV with the column density is fixed at the galactic value towards this globular cluster for all observations. Models to neutron star X-ray binary spectra can be degenerate \citep{lin07}, especially over a short energy range, such as covered by the HEG.  The model we adopt is fiducial and reflects a physically-motivated scenario consisting of thermal and non-thermal emission.  It allows us to characterize the source flux well because it provides a good fit to the spectrum.  Importantly, it is a simple continuum model and so is easily reproducible, but, due to modeling degeneracies the specific parameter values derived should be regarded with caution.  The fitted spectral parameters are given in Table \ref{tab:fluxes}.

We determine the 0.5 - 10 keV source flux, by extending a dummy response to lower and higher energies.  We determine the uncertainty in the flux by propagating the errors on the individual parameters.  The calculated fluxes are given in Table \ref{tab:fluxes}.  We note that the uncertainties in the fluxes may be dominated by calibration uncertainties, and could be closer to around 6\% (the difference in flux we find between fitting to the MEG and HEG) than those we determine from spectral fitting. 

\begin{figure}
\centering
\includegraphics[width=8cm]{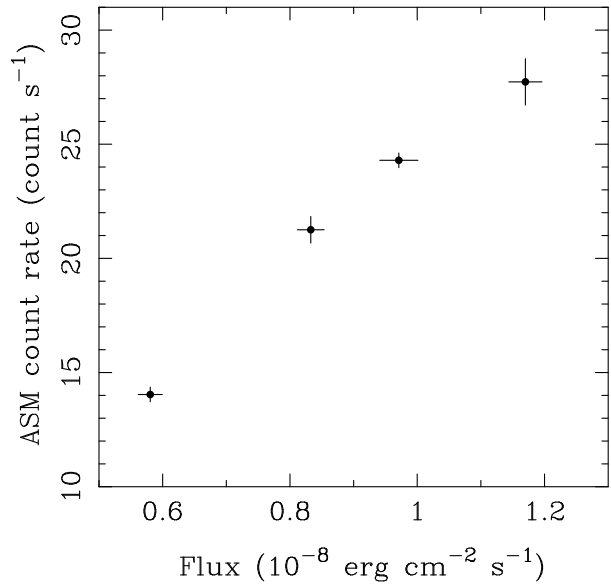}
\caption{0.5-10 keV unabsorbed source flux determined from fitting to the HEG spectra is shown versus the 1-day average {\it RXTE} All-Sky Monitor count rate on the day of each observation.}
\label{fig:asm}
\end{figure}

We check whether pileup alters the determined fluxes substantially by comparing the calculated source fluxes with the 1-day average {\it RXTE} All-Sky Monitor (ASM) count rate for the day of each {\it Chandra} observation (see Tab. \ref{tab:fluxes}).  Comparing the source flux against the ASM count rate, they  are seen to be correlated, with no large offsets (see Fig. \ref{fig:asm}).

To look for variability of the absorption lines between each epoch, in Fig. \ref{fig:ew} we plot the measured equivalent width (or upper limit, if the line is not detected) for each observation vs the 0.5-10 keV unabsorbed source flux (see Table \ref{tab:fluxes}).  The errorbars indicate 1$\sigma$ uncertainties in the equivalent width.  The dashed lines indicates the weighted mean (weighted by the uncertainties) of the equivalent widths of the detected lines.  Generally, the equivalent widths of the lines are consistent between the observations. However, we note that the \ion{O}{8} line does seem to display variability, although the variability is not highly significant and does not appear to be correlated with the source flux (see Fig.~\ref{fig:ew}).  The observed wavelength of the \ion{O}{8} line is also found to be consistent with the theoretical wavelength and not blueshifted.

We calculate the $\chi^2$ value for a fit of the equivalent widths for the \ion{O}{8} line to their weighted mean to assess the significance of this variability.  For the non-detection in ObsId 1021, we use the equivalent width determined from fitting a Gaussian fixed at the wavelength of the line in the calculation.  This equivalent width is found to $2.9\pm2.9$ m\AA. We get $\chi^2 = 14.1$ (3 degrees of freedom) when fitting the equivalent widths to their weighted mean.  For the hypothesis that the equivalent widths are constant, this corresponds to a probability for achieving a higher $\chi^2$ value of 0.003.  However, we note that we have searched 5 lines to find variability of this level, and thus this increases the probability to 0.015.  This is equivalent to the line being variable at the $2.4\sigma$ level of confidence.  Thus, while it is not a highly significant result, low-level variability cannot be ruled out for this line.

In comparing equivalent width measurements from multiple epochs, it is important to consider whether any systematic effects can cause variability in the lines.  Importantly, neither the first two observations (ObsId 1021 and 1022) were operated in continuous-clocking mode, and thus both are affected by pileup.  It is not clear how pileup alters equivalent width measurements, though we note that as the \ion{O}{7} and \ion{O}{8} lines are close together in wavelength they should be affected similarly.  Any effect must be fairly small, as the \ion{O}{8} line is seen to vary whereas the \ion{O}{7} is not.

\section{Discussion}
We have analyzed four {\it Chandra} HETGS observations of the neutron star low-mass X-ray binary 4U~1820$-$30 and detect a variety of absorption lines in the spectra.  The first two archival observations suffer from pile-up, and thus measurements of lines from these observations may not be robust.  However, the two new observations presented here do not suffer from this problem and have an increased sensitivity compared to the previous observations, allowing for a significantly improved study of the absorption lines in this system.

To investigate the nature of these lines, we compare the line equivalent widths between the observations.  If the line is unsaturated and on the linear part of the curve of growth, then the equivalent width is a direct measure of the absorbing column.  If the equivalent width does not vary between observations, this would therefore indicate that the absorbing column has remained unchanged between the observations, as would be expected for absorption lines associated with the ISM.  The majority of the absorption lines present in 4U~1820$-$30 are consistent with remaining constant between the observations, suggesting they are interstellar in origin.  However, if the lines are saturated, then even if the column density changes significantly (for instance in  response to changes in the source) the equivalent width would remain more or less constant. Similarly, in the saturation regime even if the ionizing flux changes by a factor of 2 (as observed between ObsId 6633 or 6634) the equivalent width and ion relative abundance fractions may not change by a similar amount.  However, even though we do not observe the $\alpha - \beta - \gamma$ sequence for each line, upper limits on $\beta/\alpha$ indicate that the lines are not highly saturated.

\begin{figure}
\centering
\includegraphics[width=8cm]{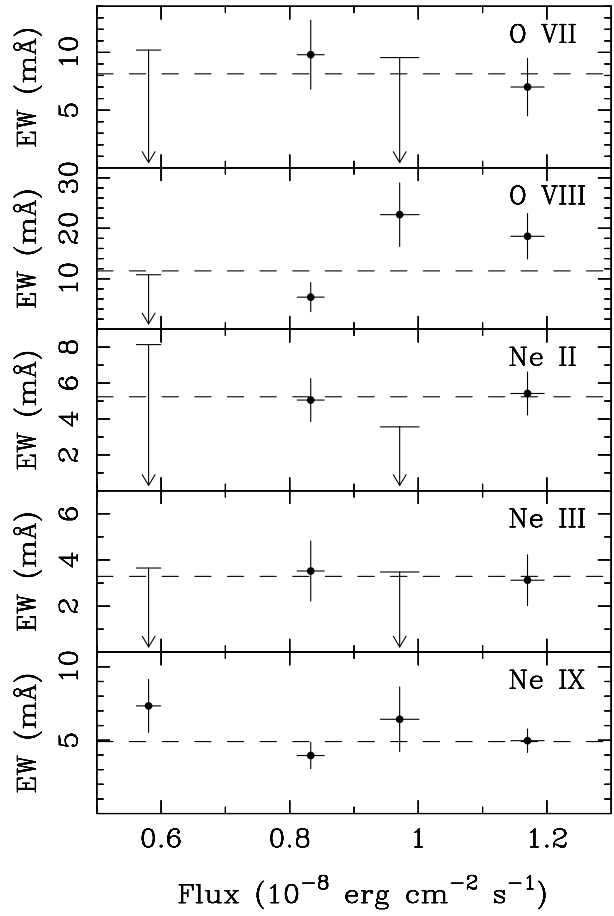}
\caption{Equivalent widths (m\AA) of the various absorption lines in 4U~1820$-$30 versus 0.5-10 keV unabsorbed source flux in units of $10^{-8}$ erg cm$^{-2}$ s$^{-1}$ . Dashed lines indicate the weighted mean of the measured equivalent widths.  Arrows indicate 95\% confidence upper limits and errorbars indicate $1-\sigma$ uncertainties.}
\label{fig:ew}
\end{figure}

We find that there may be a low level of variability (2.4-$\sigma$ significance) in the \ion{O}{8} Ly$\alpha$ line, which may indicate it is associated with the source.  If the absorption is local to the source, this variability could be, for example, due to changes in mass accretion rate leading to changes in mass outflow.  If one naively assumes that the continuum flux traces the mass accretion rate one would therefore see correlated variability between the source flux and the line equivalent width.  Alternatively, if the properties of the absorbing column are approximately constant and the ionizing flux changes, the fraction of oxygen in \ion{O}{8} should change, which could lead to correlated variability. However, there appears to be no such correlation observed (see Fig.~\ref{fig:ew}) though more observations are required to be conclusive.  

Additionally, if the gas in the system responds to changes in the ionizing flux more slowly than the inherent changes in the ionizing flux (i.e. if the recombination time is long), this would have the effect of greatly dampening the amplitude of variations seen in the absorption relative to the incident flux. For \ion{O}{8} originating within the system, the size of the absorbing region is $<10^{11}$ cm.  The column density we observe to be about $10^{20}$ cm$^{-2}$, thus the density is $n_e \sim$10$^{9}$ cm$^{-3}$.  The recombination rate coefficient for \ion{O}{8} is $\alpha = 1.5\times10^{-12}$ cm$^3$ s$^{-1}$ at 10$^6$ K, so the recombination timescale = $1/(n_e \alpha)$ $\sim$700 s.  This is significantly shorter than the timescale between the observations and thus should not be a factor.

If the absorption lines were associated with a disk wind, then blueshifts would be expected. We find that the wavelength of the absorption lines are consistent with their rest wavelengths (after accounting for the absolute instrumental calibration uncertainty of 0.05\%), as expected for interstellar lines.  However, we note that in the dipping X-ray binaries, absorption lines that are local to the source are not seen to be blueshifted and are thought to be associated with the accretion disk corona rather than an accretion disk wind.  Similarly, here the lack of blueshifts could just be explained by absorption from an accretion disk corona \citep{diaztrigo06}. 

To test both possible origins for the lines, we first calculated the equivalent hydrogen column densities (implied by the observed equivalent widths) for a variety of Doppler parameters, and find that they are consistent with the interstellar origin for the lines. Additionally we perform photoionization modeling of absorption by material local to the system.  We find that photoionization can reproduce the observed spectra well.  However, to do this requires a very low filling factor, but this could possibly be explained by absorption from dense blobs.  We conclude that while the \ion{Ne}{2} and \ion{Ne}{3} lines are produced in the ISM, the origin of the \ion{Ne}{9}, \ion{O}{7} and \ion{O}{8} lines remains unclear, with both the interstellar and local absorption remaining possible.  The key to determining the origin of these lines may lie in variability, thus additional sensitive observations of 4U~1820$-$30 are needed to confirm variability in the \ion{O}{8} line, and search for low-level variability in other lines.

\acknowledgements
We thank an anonymous referee for helpful suggestions that improved the paper.  We acknowledge B. Otte for useful discussions on observations of \ion{O}{6}. We thank the {\it RXTE}/ASM team for the quick-look results used in this paper.  JMM gratefully acknowledges support from Chandra.

\bibliographystyle{apj}
\bibliography{apj-jour,4u1820}

\begin{thebibliography}{34}
\expandafter\ifx\csname natexlab\endcsname\relax\def\natexlab#1{#1}\fi

\bibitem[{{Arnaud}(1996)}]{arnaud96}
{Arnaud}, K.~A. 1996, in ASP Conf. Ser. 101: Astronomical Data Analysis
  Software and Systems V, 17

\bibitem[{{Behar} \& {Netzer}(2002)}]{behar_netzer02}
{Behar}, E. \& {Netzer}, H. 2002, \apj, 570, 165

\bibitem[{{Bohlin} {et~al.}(1978){Bohlin}, {Savage}, \& {Drake}}]{bohlin78}
{Bohlin}, R.~C., {Savage}, B.~D., \& {Drake}, J.~F. 1978, \apj, 224, 132

\bibitem[{{Boirin} {et~al.}(2005){Boirin}, {M{\'e}ndez}, {D{\'{\i}}az Trigo},
  {Parmar}, \& {Kaastra}}]{boirin05}
{Boirin}, L., {M{\'e}ndez}, M., {D{\'{\i}}az Trigo}, M., {Parmar}, A.~N., \&
  {Kaastra}, J.~S. 2005, \aap, 436, 195

\bibitem[{{Church} {et~al.}(2005){Church}, {Reed}, {Dotani},
  {Ba{\l}uci{\'n}ska-Church}, \& {Smale}}]{church05}
{Church}, M.~J., {Reed}, D., {Dotani}, T., {Ba{\l}uci{\'n}ska-Church}, M., \&
  {Smale}, A.~P. 2005, \mnras, 359, 1336

\bibitem[{{Cottam} {et~al.}(2001{\natexlab{a}}){Cottam}, {Kahn}, {Brinkman},
  {den Herder}, \& {Erd}}]{cottam01_0748}
{Cottam}, J., {Kahn}, S.~M., {Brinkman}, A.~C., {den Herder}, J.~W., \& {Erd},
  C. 2001{\natexlab{a}}, \aap, 365, L277

\bibitem[{{Cottam} {et~al.}(2001{\natexlab{b}}){Cottam}, {Sako}, {Kahn},
  {Paerels}, \& {Liedahl}}]{cottam01_1822}
{Cottam}, J., {Sako}, M., {Kahn}, S.~M., {Paerels}, F., \& {Liedahl}, D.~A.
  2001{\natexlab{b}}, \apjl, 557, L101

\bibitem[{{Crenshaw} {et~al.}(1999){Crenshaw}, {Kraemer}, {Boggess}, {Maran},
  {Mushotzky}, \& {Wu}}]{crenshaw99}
{Crenshaw}, D.~M., {Kraemer}, S.~B., {Boggess}, A., {Maran}, S.~P.,
  {Mushotzky}, R.~F., \& {Wu}, C.-C. 1999, \apj, 516, 750

\bibitem[{{D{\'{\i}}az Trigo} {et~al.}(2006){D{\'{\i}}az Trigo}, {Parmar},
  {Boirin}, {M{\'e}ndez}, \& {Kaastra}}]{diaztrigo06}
{D{\'{\i}}az Trigo}, M., {Parmar}, A.~N., {Boirin}, L., {M{\'e}ndez}, M., \&
  {Kaastra}, J.~S. 2006, \aap, 445, 179

\bibitem[{{Dickey} \& {Lockman}(1990)}]{dickeylockman90}
{Dickey}, J.~M. \& {Lockman}, F.~J. 1990, \araa, 28, 215

\bibitem[{{Futamoto} {et~al.}(2004){Futamoto}, {Mitsuda}, {Takei}, {Fujimoto},
  \& {Yamasaki}}]{futamoto04}
{Futamoto}, K., {Mitsuda}, K., {Takei}, Y., {Fujimoto}, R., \& {Yamasaki},
  N.~Y. 2004, \apj, 605, 793

\bibitem[{{Juett} {et~al.}(2004){Juett}, {Schulz}, \& {Chakrabarty}}]{juett04}
{Juett}, A.~M., {Schulz}, N.~S., \& {Chakrabarty}, D. 2004, \apj, 612, 308

\bibitem[{{Juett} {et~al.}(2006){Juett}, {Schulz}, {Chakrabarty}, \&
  {Gorczyca}}]{juett06}
{Juett}, A.~M., {Schulz}, N.~S., {Chakrabarty}, D., \& {Gorczyca}, T.~W. 2006,
  \apj, 648, 1066

\bibitem[{{Kaastra} {et~al.}(2000){Kaastra}, {Mewe}, {Liedahl}, {Komossa}, \&
  {Brinkman}}]{kaastra00}
{Kaastra}, J.~S., {Mewe}, R., {Liedahl}, D.~A., {Komossa}, S., \& {Brinkman},
  A.~C. 2000, \aap, 354, L83

\bibitem[{{Kallman} \& {Bautista}(2001)}]{kallman01}
{Kallman}, T. \& {Bautista}, M. 2001, \apjs, 133, 221

\bibitem[{{Kaspi} {et~al.}(2000){Kaspi}, {Brandt}, {Netzer}, {Sambruna},
  {Chartas}, {Garmire}, \& {Nousek}}]{kaspi00}
{Kaspi}, S., {Brandt}, W.~N., {Netzer}, H., {Sambruna}, R., {Chartas}, G.,
  {Garmire}, G.~P., \& {Nousek}, J.~A. 2000, \apjl, 535, L17

\bibitem[{{Kuulkers} {et~al.}(2003){Kuulkers}, {den Hartog}, {in't Zand},
  {Verbunt}, {Harris}, \& {Cocchi}}]{kuulkers03}
{Kuulkers}, E., {den Hartog}, P.~R., {in't Zand}, J.~J.~M., {Verbunt},
  F.~W.~M., {Harris}, W.~E., \& {Cocchi}, M. 2003, \aap, 399, 663

\bibitem[{{Lee} {et~al.}(2002){Lee}, {Reynolds}, {Remillard}, {Schulz},
  {Blackman}, \& {Fabian}}]{lee02}
{Lee}, J.~C., {Reynolds}, C.~S., {Remillard}, R., {Schulz}, N.~S., {Blackman},
  E.~G., \& {Fabian}, A.~C. 2002, \apj, 567, 1102

\bibitem[{{Lin} {et~al.}(2007){Lin}, {Remillard}, \& {Homan}}]{lin07}
{Lin}, D., {Remillard}, R.~A., \& {Homan}, J. 2007, ArXiv Astrophysics
  e-prints, astro-ph/0702089

\bibitem[{{Mazzotta} {et~al.}(1998){Mazzotta}, {Mazzitelli}, {Colafrancesco},
  \& {Vittorio}}]{mazzotta98}
{Mazzotta}, P., {Mazzitelli}, G., {Colafrancesco}, S., \& {Vittorio}, N. 1998,
  \aaps, 133, 403

\bibitem[{{Miller} {et~al.}(2006{\natexlab{a}}){Miller}, {Raymond}, {Fabian},
  {Steeghs}, {Homan}, {Reynolds}, {van der Klis}, \&
  {Wijnands}}]{miller06_nature}
{Miller}, J.~M., {Raymond}, J., {Fabian}, A., {Steeghs}, D., {Homan}, J.,
  {Reynolds}, C., {van der Klis}, M., \& {Wijnands}, R. 2006{\natexlab{a}},
  \nat, 441, 953

\bibitem[{{Miller} {et~al.}(2004){Miller}, {Raymond}, {Fabian}, {Homan},
  {Nowak}, {Wijnands}, {van der Klis}, {Belloni}, {Tomsick}, {Smith},
  {Charles}, \& {Lewin}}]{miller04_gx339}
{Miller}, J.~M., {Raymond}, J., {Fabian}, A.~C., {Homan}, J., {Nowak}, M.~A.,
  {Wijnands}, R., {van der Klis}, M., {Belloni}, T., {Tomsick}, J.~A., {Smith},
  D.~M., {Charles}, P.~A., \& {Lewin}, W.~H.~G. 2004, \apj, 601, 450

\bibitem[{{Miller} {et~al.}(2006{\natexlab{b}}){Miller}, {Raymond}, {Homan},
  {Fabian}, {Steeghs}, {Wijnands}, {Rupen}, {Charles}, {van der Klis}, \&
  {Lewin}}]{miller06_H1743}
{Miller}, J.~M., {Raymond}, J., {Homan}, J., {Fabian}, A.~C., {Steeghs}, D.,
  {Wijnands}, R., {Rupen}, M., {Charles}, P., {van der Klis}, M., \& {Lewin},
  W.~H.~G. 2006{\natexlab{b}}, \apj, 646, 394

\bibitem[{{Otte} \& {Dixon}(2006)}]{otte06}
{Otte}, B. \& {Dixon}, W.~V.~D. 2006, \apj, 647, 312

\bibitem[{{Parmar} {et~al.}(2002){Parmar}, {Oosterbroek}, {Boirin}, \&
  {Lumb}}]{parmar02}
{Parmar}, A.~N., {Oosterbroek}, T., {Boirin}, L., \& {Lumb}, D. 2002, \aap,
  386, 910

\bibitem[{{Schulz} {et~al.}(2001){Schulz}, {Chakrabarty}, {Marshall},
  {Canizares}, {Lee}, \& {Houck}}]{schulz01}
{Schulz}, N.~S., {Chakrabarty}, D., {Marshall}, H.~L., {Canizares}, C.~R.,
  {Lee}, J.~C., \& {Houck}, J. 2001, \apj, 563, 941

\bibitem[{{Sidoli} {et~al.}(2001){Sidoli}, {Oosterbroek}, {Parmar}, {Lumb}, \&
  {Erd}}]{sidoli01}
{Sidoli}, L., {Oosterbroek}, T., {Parmar}, A.~N., {Lumb}, D., \& {Erd}, C.
  2001, \aap, 379, 540

\bibitem[{{Spitzer}(1978)}]{spitzer78}
{Spitzer}, L. 1978, {Physical processes in the interstellar medium} (New York
  Wiley-Interscience, 1978.~333 p.)

\bibitem[{{Stella} {et~al.}(1987){Stella}, {Priedhorsky}, \&
  {White}}]{stella87}
{Stella}, L., {Priedhorsky}, W., \& {White}, N.~E. 1987, \apjl, 312, L17

\bibitem[{{Ueda} {et~al.}(2004){Ueda}, {Murakami}, {Yamaoka}, {Dotani}, \&
  {Ebisawa}}]{ueda04}
{Ueda}, Y., {Murakami}, H., {Yamaoka}, K., {Dotani}, T., \& {Ebisawa}, K. 2004,
  \apj, 609, 325

\bibitem[{{Verner} {et~al.}(1996){Verner}, {Verner}, \& {Ferland}}]{verner96}
{Verner}, D.~A., {Verner}, E.~M., \& {Ferland}, G.~J. 1996, Atomic Data and
  Nuclear Data Tables, 64, 1

\bibitem[{{Wilms} {et~al.}(2000){Wilms}, {Allen}, \& {McCray}}]{wilms00}
{Wilms}, J., {Allen}, A., \& {McCray}, R. 2000, \apj, 542, 914

\bibitem[{{Yao} \& {Wang}(2005)}]{yao_wang05}
{Yao}, Y. \& {Wang}, Q.~D. 2005, \apj, 624, 751

\bibitem[{{Yao} \& {Wang}(2006)}]{yao_wang06}
---. 2006, \apj, 641, 930

\end{thebibliography}

\clearpage

\begin{deluxetable}{cccc}
\tablecolumns{4} 
\tablewidth{0pc}
\tablecaption{{\it Chandra} observations of 4U 1820$-$30\label{tab:obs}}
\tablehead{ObsID & Observation Date & Grating/Detector & Exposure (ks)}
\startdata
98 & 2000 Mar 10 & LETG/HRC & 15.12\\
1021 & 2001 Jul 21 & HETG/ACIS & 9.70\\
1022 & 2001 Sep 12 & HETG/ACIS & 10.89\\
6633 & 2006 Aug 12  & HETG/ACIS & 25.18\\
6634 & 2006 Oct 20 & HETG/ACIS & 25.13
\enddata
\tablecomments{ObsID 6633 and 6634 are the new observations presented in this work.}
\end{deluxetable}

\begin{deluxetable*}{lccccccc}
\tabletypesize{\footnotesize}
\tablecolumns{8} 
\tablewidth{0pc}
\tablecaption{Absorption line properties in 4U 1820$-$30}
\tablehead{ObsID & Ion & Transition & Theoretical  & Observed & Equivalent width  & FWHM  & Flux  \\
 & & &(\AA) & (\AA) & (m\AA) & (m\AA) & (10$^{-4}$ photons cm$^{-2}$ s$^{-1}$) }
\startdata
1021&\ion{Ne}{9}& 1s$^2-$1s2p & 13.447 & $13.455\pm0.003$  & $7.3\pm1.8$  & $3.8^{+7.3}_{-3.5}$  & $4.3\pm1.0$ \\
 & \ion{Ne}{3}  &1s$-$2p      & 14.508 &  & $<3.7$ & & $<2.1$\\
 & \ion{Ne}{2}  &1s$-$2p      & 14.608 &  & $<8.1$ & & $<4.7$\\ 
 & \ion{O}{7}   & 1s$^2-$1s3p & 18.629 &  & $<10.2$ & & $<4.8$ \\
 & \ion{O}{8}   & 1s$-$2p     & 18.967  &  & $<10.8$ & & $<4.8$ \\
\hline
 1022 & \ion{Ne}{9} &1s$^2-$1s2p& 13.447 & $13.448\pm0.005$  & $6.4\pm2.2$ & $19.3\pm10.9$ & $5.2\pm1.8$ \\
 & \ion{Ne}{3} &1s$-$2p& 14.508 &  & $<3.5$ & & $<2.7$\\
 & \ion{Ne}{2} &1s$-$2p& 14.608 &  & $<3.6$ & & $<2.8$\\ 
 & \ion{O}{7}  &1s$^2-$1s3p& 18.629 &  & $<6.6$ & & $<6.0$\\
 & \ion{O}{8} &1s$-$2p& 18.967  & $18.966\pm0.007$ & $22.7\pm6.3$ & $33.4\pm14.1$ & $1.4\pm0.4$ \\
\hline
 6633 & \ion{Ne}{9} &1s$^2-$1s2p& 13.447 &  $13.450\pm0.002$ & $4.0\pm0.9$ & $1.4^{+9.7}_{-1.4}$ & $2.9\pm0.7$\\
 & \ion{Ne}{3} &1s$-$2p& 14.508 & $14.505\pm0.006$ & $3.5\pm1.3$ & $18.6\pm13.5$ & $2.5\pm 0.9$\\
 & \ion{Ne}{2} &1s$-$2p& 14.608 & $14.611\pm0.002$ & $5.1\pm1.2$ & $0.07^{+11.6}_{-0.07}$ & $3.5\pm0.8$\\
 & \ion{O}{7}  &1s$^2-$1s3p& 18.629 & $18.634\pm0.006$ & $9.8\pm3.0$ & $24.0\pm11.0$ & $6.3\pm2.0$\\
 & \ion{O}{8} &1s$-$2p& 18.967  & $18.972\pm0.006$ & $6.4\pm2.9$ & $7.1^{+21.2}_{-7.1}$  & $4.1\pm1.8$ \\
\hline
 6634 & \ion{Ne}{9} &1s$^2-$1s2p& 13.447 & $13.445\pm0.002$ & $5.0\pm0.8$ & $1.4^{+8.6}_{-1.4}$ & $4.9\pm0.8$ \\
 & \ion{Ne}{3} &1s$-$2p& 14.508 & $14.508\pm0.003$ & $3.1\pm1.1$ & $3.2^{+13.0}_{-3.2}$ & $3.0\pm1.0$\\
 & \ion{Ne}{2} &1s$-$2p& 14.608 & $14.602\pm0.005$ & $5.4\pm1.2$ & $27.8\pm12.2$ & $5.1\pm1.2$\\
 & \ion{O}{7}  &1s$^2-$1s3p& 18.629 & $18.625^{+0.007}_{-0.001}$ & $7.0\pm2.5$ & $0.02^{+14.4}_{-0.02}$ & $6.1\pm2.3$\\
 & \ion{O}{8} &1s$-$2p& 18.967 & $18.965\pm0.006$ & $18.4\pm4.5$ & $45.9\pm12.6$ & $16.0\pm4.0$ 
\enddata
\label{tab:lines}
\tablecomments{See text for details on spectral fitting procedure.  Theoretical line wavelengths for \ion{Ne}{9}, \ion{O}{7}, and \ion{O}{8} are from \citet*{verner96}.  The line wavelengths for \ion{Ne}{2} and \ion{Ne}{3} are the best observationally determined values from \citet{juett06}.  Uncertainties and upper limits are 1$\sigma$ confidence limits.  Where the line was not detected and we determine upper limits, the wavelength was fixed at the theoretical value, and the width of the Gaussian fixed below the instrument resolution.}
\end{deluxetable*}

\begin{deluxetable*}{llcccccccc}
\tabletypesize{\footnotesize}
\tablecolumns{10} 
\tablewidth{0pc}
\tablecaption{Line column densities for various Doppler parameters}
\tablehead{ObsID & Line & \multicolumn{2}{c}{Linear curve of growth} & \multicolumn{2}{c}{b = 50 km s$^{-1}$}  & \multicolumn{2}{c}{b = 100 km s$^{-1}$} & \multicolumn{2}{c}{b = 200 km s$^{-1}$} \\
& & $N_{ion}$ & $N_{H}$ & $N_{ion}$ & $N_{H}$ & $N_{ion}$ & $N_{H}$ & $N_{ion}$ & $N_{H}$ \\
& & $(10^{16}$ cm$^{-2}$) & $(10^{20}$ cm$^{-2}$) & $(10^{16}$ cm$^{-2}$) & $(10^{20}$ cm$^{-2}$)  & $(10^{16}$ cm$^{-2}$) & $(10^{20}$ cm$^{-2}$) & $(10^{16}$ cm$^{-2}$) & $(10^{20}$ cm$^{-2}$) 
}
\startdata
6633 & \ion{Ne}{9} (13.447\AA) & 0.3 & 0.4 & 0.6 & 0.7 & 0.4 & 0.5 & 0.4 & 0.4\\
     & \ion{Ne}{3} (14.508\AA) & 1.8 & 7.1 & 2.6 & 12.5 & 2.1 & 8.9 & 1.9 & 7.8\\
     & \ion{Ne}{2} (14.608\AA) & 4.4 & 7.1 & 8.3 & 12.6 & 5.6 & 8.9 & 4.9 & 7.8\\
     & \ion{O}{7} (18.629\AA)  & 2.2 & 0.5 & 9.7 & 2.1 & 3.3 & 0.8 & 2.6 & 0.6\\
     & \ion{O}{8} (18.967\AA)  & 0.5 & 0.6 & 0.7 & 2.1 & 0.6 & 0.9 & 0.5 & 0.7\\
\hline
6634 & \ion{Ne}{9} (13.447\AA) & 0.4 & 0.5 & 0.9 & 1.1 & 0.6 & 0.7 & 0.5 & 0.6\\
     & \ion{Ne}{3} (14.508\AA) & 1.6 & 7.1 & 2.2 & 13.4 & 1.8 & 9.0  & 1.7 & 7.9 \\
     & \ion{Ne}{2} (14.608\AA) & 4.6 & 7.1 & 9.4 & 13.4 & 6.0 & 9.1 & 5.2 & 8.0\\
     & \ion{O}{7} (18.629\AA)  & 1.6 & 0.6 & 3.2 & 1.3 & 2.1 & 0.8 & 1.8 & 0.7\\
     & \ion{O}{8} (18.967\AA)  & 1.4 & 0.8 & 6.4 & 3.2 & 2.9 & 1.5 & 1.9 & 1.0\\
\enddata
\label{tab:densities}
\tablecomments{See text for details on determining column densities.  We only determine column densities for ObsIDs 6633 and 6634.  The equivalent hydrogen column density ($N_H$) is determined using the ISM abundances from \citet{wilms00}.}
\end{deluxetable*}

\begin{deluxetable*}{lccccccc}
\tabletypesize{\footnotesize}
\tablecolumns{2} 
\tablewidth{0pc}
\tablecaption{Spectral fits to 4U 1820$-$30}
\tablehead{ObsID & \multicolumn{2}{c}{Blackbody} & \multicolumn{2}{c}{Power-law} & $\chi^2_\nu$ & $0.5-10$ keV unabsorbed flux & {\it RXTE}/ASM count rate\\
 & kT (keV) & Normalization & Spectral index & Normalization & & (10$^{-9}$ erg cm$^{-2}$ s$^{-1}$) & (counts s$^{-1}$)}
\startdata
1021 & 0.72$\pm 0.07$ & $(3.5 \pm 0.6) \times 10^{-3}$ & $1.67\pm0.02$	& $0.85\pm0.02$ & 0.87 & $5.8 \pm  0.2$ & $14.0 \pm 0.3$\\
1022 & $0.96^{+0.15}_{-0.11}$ & $(4.5\pm1.0)\times10^{-3}$ & $1.46\pm0.02$	& $1.13\pm0.02$ & 0.88 & $9.7 \pm  0.3$ & $24.3 \pm 0.3$\\
6633 & $1.16\pm0.02$ & $(2.1\pm0.1)\times10^{-2}$ & $1.76\pm0.03$ & $1.12\pm0.01$ & 0.94 & $8.3 \pm  0.2$ & $21.3 \pm 0.6$\\
6634 & $1.20\pm0.02$ & $(3.2\pm0.2)\times10^{-2}$ & $1.74\pm0.02$ & $1.50\pm0.01$ & 1.24 & $11.7 \pm 0.3$ & $27.7 \pm 1.0$
\enddata
\label{tab:fluxes}
\tablecomments{In all fits the column density was fixed at the cluster value of $N_H = 1.5 \times 10^{21}$ cm$^{-2}$.  We used the XSPEC blackbody model named bbody, where the normalization is defined as L39/(D10)$^2$, where L39 is the luminosity in units of 10$^{39}$ erg s$^{-1}$ and D10 the distance to the source in units of 10 kpc. The power-law normalization is defined at 1 keV, with units of photons keV$^{-1}$ cm$^{-2}$ s$^{-1}$. The spectral fits are to the HEG 1st order spectrum (see text for details).  For comparison with the source flux, the {\it RXTE}/ASM count rate for 4U 1820$-$30 is also given.}
\end{deluxetable*}

\end{document}